\documentclass[aps,prlprint, twocolumn,superscriptaddress,amsmath,amssymb,showpacs]{revtex4}

\usepackage{colortbl}
\usepackage{amsmath}
\usepackage{amssymb}
\usepackage{gensymb}

\usepackage[OT2,OT1]{fontenc}
\newcommand\cyr
{
\renewcommand\rmdefault{wncyr}
\renewcommand\sfdefault{wncyss}
\renewcommand\encodingdefault{OT2}
\normalfont
\selectfont
}
\DeclareTextFontCommand{\textcyr}{\cyr}

\usepackage{bm}

	\usepackage{amsmath}
	\usepackage{makeidx}
	\usepackage{amsfonts}
	\usepackage[ansinew]{inputenc}
	\usepackage[usenames,dvipsnames]{pstricks}
	\usepackage{subfigure}
	\usepackage{epsfig}
	\usepackage{pst-grad} 
	\usepackage{pst-plot} 
	\usepackage[colorlinks,hyperindex]{hyperref}
	\usepackage{tikz}
	\hypersetup
	{
		colorlinks,%
		citecolor=black,%
		linkcolor=black,%
		urlcolor=black,%
	}

\usepackage{gensymb}

\newcommand{\beq}{\begin{equation}}
\newcommand{\eeq}{\end{equation}}
\newcommand{\vecp}{{\bf p}}

\newcommand{\vecj}{{\bf j}}
\newcommand{\vecB}{{\bf B}}
\newcommand{\vecA}{{{\bf A}}}

\makeindex

\begin{document}

\title{Laser electron acceleration on curved surfaces}
\author{Ph~ Korneev} 
\email{korneev@theor.mephi.ru}
\affiliation{National Research Nuclear University MEPhI, Moscow, Russian Federation}
\affiliation{P.N.~Lebedev Physical Institute of RAS, Moscow, Russian Federation} 
\author{ Y.~Abe}
\affiliation{Institute of Laser Engineering, Osaka University, Japan}
\author{  K.-F.-F.~Law}
\affiliation{Institute of Laser Engineering, Osaka University, Japan}
\author{S.~G.~Bochkarev}
\affiliation{P.~N.~Lebedev Physical Institute of RAS, Moscow, Russian Federation} 
\author{ S.~Fujioka}
\affiliation{Institute of Laser Engineering, Osaka University, Japan}
\author{  S.~Kojima}
\affiliation{Institute of Laser Engineering, Osaka University, Japan}
\author{  S.-H.~Lee}
\affiliation{Institute of Laser Engineering, Osaka University, Japan}
\author{ S.~Sakata}
\affiliation{Institute of Laser Engineering, Osaka University, Japan}
\author{  K.~Matsuo}
\affiliation{Institute of Laser Engineering, Osaka University, Japan}
\author{  A.~Oshima}
\affiliation{Institute of Laser Engineering, Osaka University, Japan}
\author{  A.~Morace}
\affiliation{Institute of Laser Engineering, Osaka University, Japan}
\author{ Y.~Arikawa}
\affiliation{Institute of Laser Engineering, Osaka University, Japan}
\author{  A.~Yogo}
\affiliation{Institute of Laser Engineering, Osaka University, Japan}
\author{  M.~Nakai}
\affiliation{Institute of Laser Engineering, Osaka University, Japan}
\author{ T.~Norimatsu}
\affiliation{Institute of Laser Engineering, Osaka University, Japan}
\author{E.~d'Humi\`eres}
\affiliation{University of Bordeaux, CNRS, CEA, CELIA, 33405 Talence, France}
\author{J.J.~Santos}
\affiliation{University of Bordeaux, CNRS, CEA, CELIA, 33405 Talence, France}
\author{ K.~Kondo}
\affiliation{National Institute for Quantum and Radiological Science and Technology, Japan}
\author{ A.~Sunahara}
\affiliation{Purdue University, USA}
\author{V.Yu.~Bychenkov}
\affiliation{P.~N.~Lebedev Physical Institute of RAS, Moscow, Russian Federation} 
\author{S. Gus'kov}
\affiliation{P.~N.~Lebedev Physical Institute of RAS, Moscow, Russian Federation} 
\affiliation{National Research Nuclear University MEPhI, Moscow, Russian Federation}
\author{V. Tikhonchuk}
\affiliation{University of Bordeaux, CNRS, CEA, CELIA, 33405 Talence, France}



\begin{abstract}
{Electron acceleration by relativistically intense laser beam propagating along a curved surface allows to split softly the accelerated electron bunch and the laser beam. The presence of a curved surface allows to switch an adiabatic invariant of electrons in the wave instantly leaving the gained energy to the particles. The efficient acceleration is provided by the presence of strong transient quasistationary fields in the interaction region and a long efficient acceleration length. The curvature of the surface allows to select the accelerated particles and provides their narrow angular distribution. The mechanism at work is explicitly demonstrated in theoretical models and experiments.        
}
\end{abstract}


\maketitle



In laser-matter interactions, electrons are the first to be affected by strong laser irradiation, absorbing energy, forming currents, and inducing fields in the interaction process. They are accelerated by intense laser pulse \cite{Kruer-pof85, Wilks-prl92} mainly along the electric field component at sub-relativistic intensities and along the Poynting vector in relativistic laser beams. The latter are used conventionally for producing dense energetic electron bunches \cite{Pukhov-pop98, Beg-pop97, Malka-pre02}, which are interesting for many applications, including laser ion acceleration \cite{Wilks-pop01} and fast ignition in the context of inertial confinement fusion \cite{Tabak-pop94}. One of the issues to be solved in the problem of direct relativistic electron acceleration is the breakdown of the ${\vecp_e} - e/c {\vecA}$ invariant \cite{Lawson-report75,Woodward-ee46} (${\vecp_e}$ is the electron momentum, $e$ is the elementary electric charge, $c$ is the light velocity and ${\vecA}$ is the laser vector potential), which is conserved in a plane electromagnetic wave. Then, the energy gained by an electron would remain with it when it leaves the accelerating laser field. The well-known approach is to introduce a strong perturbation into the electron motion, but still the deceleration at the rear part of the laser pulse may substantially reduce the gained energy. 

In this context, electron acceleration at a grazing laser incidence on solid foils has been considered. Nakamura et al. showed that if the laser beam incident angle is larger than a critical value, electrons can be transported and accelerated to high energies along the surface by quasistatic electromagnetic fields \cite{Nakamura-prl04}. Chen et al. showed that these energetic electrons perform betatron oscillations and can be accelerated significantly in the reflected laser field if the betatron frequency is close to the laser frequency in the particle frame \cite{Chen-oe06}. Studying the curved cone targets with a trumpet-like curvature, Kluge et al. showed that electron acceleration in this case is governed by two scenarios, related to action of longitudinal or transversal electric fields at the target surface, both of them leading to continuous electron acceleration along the inner cone wall \cite{Kluge-njp12}. Andreev et al. studied the effect of the plasma density distribution on electron acceleration and demonstrated a substantial increase of the energy and number of accelerated electrons for the grazing incidence of a subpicosecond intense laser pulse in comparison with the ponderomotive scaling \cite{Andreev-lpb14}. They also noted the importance of stochastic acceleration for large preplasma lengths. Yi et al. studied direct acceleration of electrons by a CO$_2$ laser in a curved plasma waveguide showing acceleration gradients as high as 26 GV/m for a 1.3 TW CO$_2$ laser system \cite{Yi-pr16}. An interesting effect, which manifests itself at grazing incidence of a laser beam, is the excitation of surface plasma waves, which may enhance laser absorption and produce high-energy electrons \cite{Kupersztych-pop04, Raynaud-pop07, Bigongiari-pop11}.

In this paper we study electron acceleration at a grazing incidence on curved targets in a form of a circular segment. The interaction geometry possesses attributes of the grazing incidence on a flat surface \cite{Nakamura-prl04, li-prl06}, but increases the acceleration path, induces the magnetic field convection further into the interaction region \cite{Korneev-pre15}, and offers a new simple and elegant feature of splitting a laser beam and the accelerated electron bunch. That is, with such a geometry of interaction, a curved surface works as a mirror for the laser and a guideline and energy selector for the electrons at the same time. 


%
%
%



Let us recall first the physics of interaction of a laser beam at a grazing incidence with a solid flat surface. The mechanism of the electron guiding and acceleration along the surface is explained in Refs. \cite{Nakamura-prl04, li-prl06}. The important feature is spontaneous generation of a quasistatic field structure, which is channeling the electron bunch. Its two main elements are the magnetic field of return currents and the charge separation electric field. The electrons, trapped in this quasistatic transient structure, may be efficiently accelerated by the incident laser beam and form a narrow directed bunch.

\begin{figure}{\begin{center}\includegraphics[width=0.4\textwidth]
{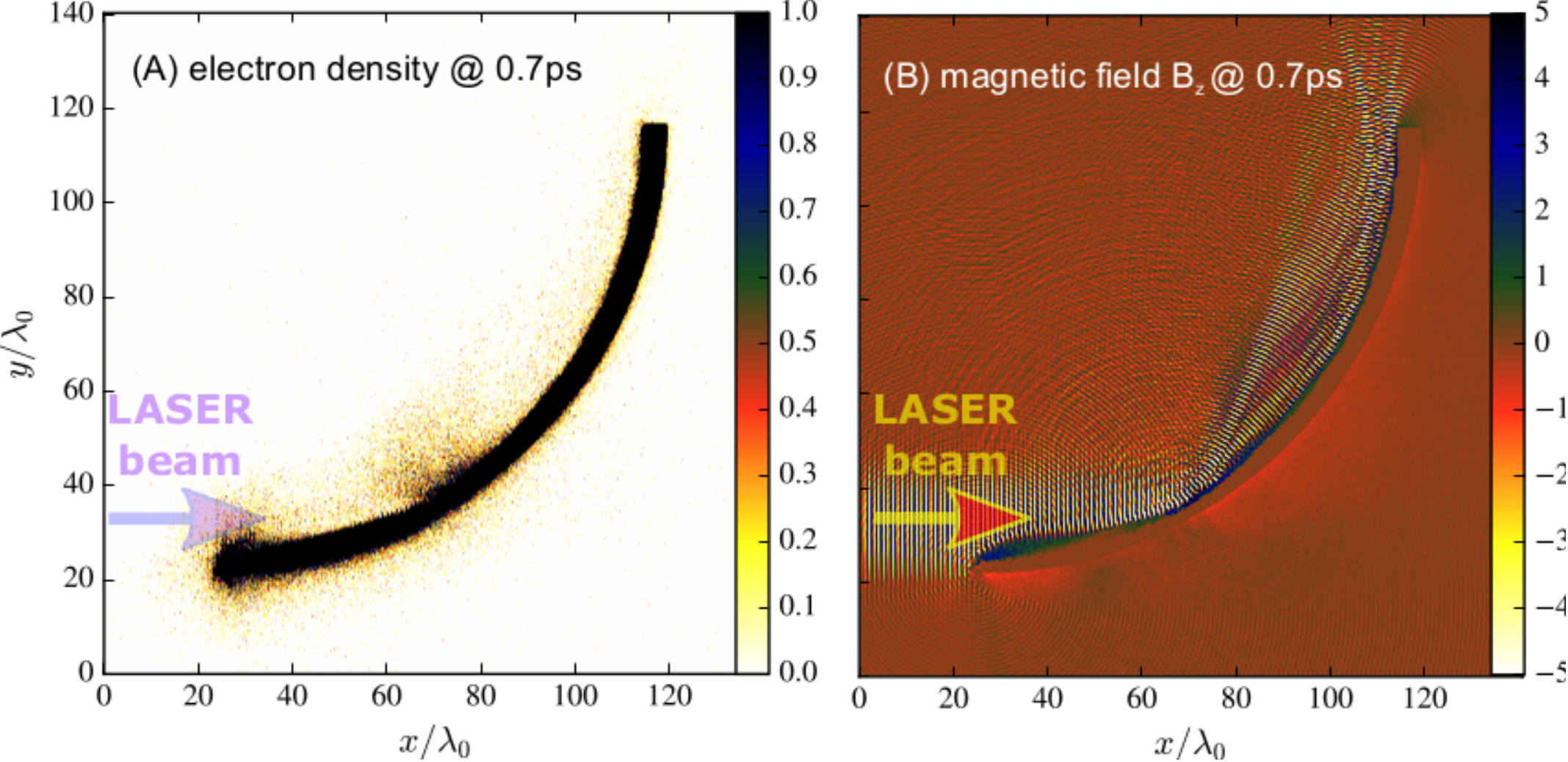}\end{center}}
\caption{Electron density (A) and $B_z$ component of the magnetic field (B) at 0.7 ps. Colorbar ranges are cut for better visibility. Mutiple reflection of the laser beam is seen in panel (B). (The electron density unit is the critical density $n_{cr}\approx 10^{21}$cm$^{-3}$, the magnetic field unit is $B_{unit}\approx10$ kT.}
\label{target}
\end{figure}

The curvature of the target brings new phenomena into play. The first one is a more developed magnetic field structure generated near the surface and further convected in vacuum with a hot expanding plasma \cite{Korneev-pre15}. 
The strong magnetic field may guide electrons even along a curved surface, repeating its geometry, if the curvature radius is not too small. At the same time, the laser beam follows the same surface due to multiple reflections. Though it is gradually depleted, for a certain length it may follow an electron trajectory, see Fig. \ref{target}, providing a much longer acceleration time than that in the case of a flat target surface. 
Another important feature of the interaction process in this geometry is the possibility of electrons to escape the interaction region through the target surface. Guided electrons may pass several reflections, unless the pitch angle $\theta\sim {p_e}_\bot/p_e$, where $p_e$ is the absolute value of the electron momentum, ${p_e}_\bot$ is its surface-normal component, becomes greater than a certain angle $\theta_0$.  

\begin{figure}{\begin{center}\includegraphics[width=0.5\textwidth]
{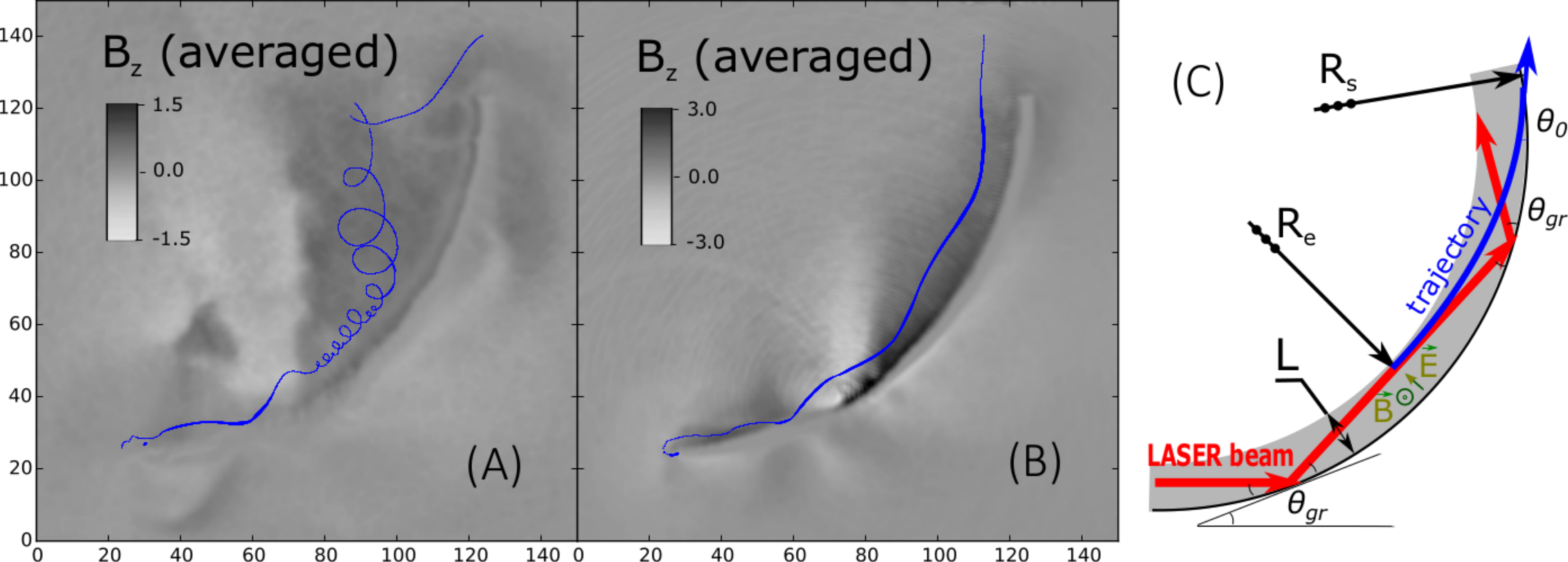}\end{center}}
\caption{Trajectories of test electrons in the PIC simulation. Panel (A): magnetized electron (at 1.3 ps); panel (B): accelerated and then decelerated (at 0.8 ps) electron. Background shows magnetic fields, averaged in space, at time moments, corresponding to the end of the shown trajectory. (The magnetic field unit is $B_{unit}\approx10$ kT, the electric field unit is $E_{unit}\approx 3$TV/m.)  Panel (C): a primitive scheme of the geometrical constraints of electron motion.}
\label{naccel}
\end{figure}

To demonstrate the discussed features, we performed 2D particle-in-cell (PIC) simulations with the code {PICLS} \cite{Sentoku-jcp08} of a p-polarized, 1 ps, $10^{19}$ W/cm$^2$ laser beam with 1 $\mu$m wavelength, irradiating a curved target at a grazing incidence. The geometry of the target and interaction may be seen in Fig. \ref{target}, where left panel (A) shows the electron density at $0.7$ ps, and right panel (B) shows the $B_z$ magnetic field component at the same time. The parameters correspond to a dimensionless amplitude $a_0=eE_0/m_ec\omega_0\approx3$, where $E_0$ and $\omega_0$ are the laser electric field amplitude and the laser frequency respectively. The target, shown in Fig.~\ref{target}, is a $\pi/2$ sector of a cylinder with 100 $\mu$m radius and a $5~\mu$m thickness. It is modelled by initially cold electron-proton plasma with $1.7\times10^{22}$  cm$^{-3}$ density. The simulation box contained $2816\times2808$ cells, with $10$ particles per cell, the resolution was $0.1$ fs in time and $30$ nm in space. Multiple reflections of the laser beam, which are seen in panel (B), are stretched along the target surface due to the beam finite focal size. This allows the beam to interact with the target electrons near the surface along all their path ``inside'' the target. 

The particle dynamics was studied with a test-particle module introduced in the PIC code. We considered the particles with the greatest instant energy during the whole interaction process, and distinguished three different scenarios. In Fig. \ref{naccel} two sample electron trajectories are shown on the background of the averaged $z$-component of the magnetic field, the line thickness
 is proportional to the particle energy. The selected electrons in Fig. \ref{naccel} gain energy from the laser field, but then lose it. The panel (A) shows the trajectory of an electron, which is strongly magnetized, at the time moment $1.3$ ps, when the laser pulse is already gone, and the electron just reached the end of the target. The panel (B) shows the trajectory of an electron, decelerated in the rear part of the laser pulse, at the time moment $0.8$ ps. 
  This electron after subsequent reflections returned its energy back to the laser beam, as it occurs in a plane wave, conserving the adiabatic invariant \cite{Lawson-report75,Woodward-ee46}.

The third scenario presents an efficient electron acceleration along the target surface. Guiding of an electron takes place, if its velocity is aligned along the surface and the radial electric force $eE_r$ deviating electron from the target surface is partially compensated by the magnetic force $ev_eB_z/c$, where $v_e$ is the electron velocity, $E_r$ and $B_z$ are the average electric radial and the magnetic $z$ field components. In fact, the net force acting on the electron, $e(E_r-B_z)$ (assuming $v_e\approx c$) should make the curvature radius of the electron trajectory $R_e\sim [e(B_z-E_r)/m_e\gamma_e c^2]^{-1}$, where $\gamma_e=1/\sqrt{1-v^2/c^2}$, to be close to the target curvature $R_s$.  
The difference between $R_e$ and $R_s$ is restricted by the condition of the electron confinement $\theta_0^2\sim2L|R_s^{-1}-R_e^{-1}|$, where $L$ is the characteristic scale of the plasma near the surface, see Fig. \ref{naccel}, panel (C).  
Then $\gamma_e\sim (e/m_e \omega_0^2)(B_z-E_r)[R_s^{-1}-\theta_{0}^2/2L]^{-1}$. A sharp dependence of $\gamma_e$ on $\theta_0$ indicates the presence of a narrow peak in the angular distribution of the most energetic electrons.

\begin{figure}{\begin{center}\includegraphics[width=0.4\textwidth]
{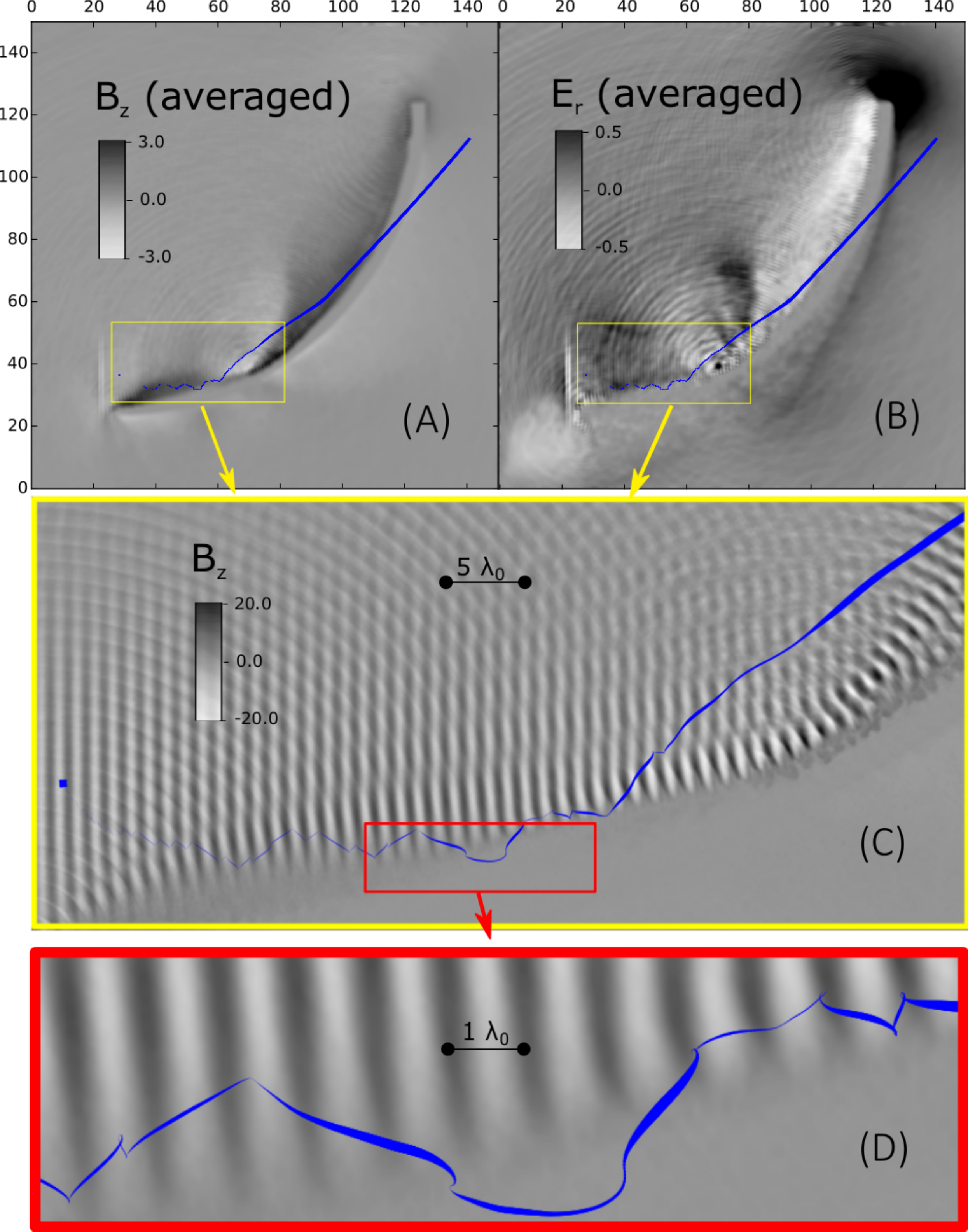}\end{center}}
\caption{Trajectory of a test electron, accelerated and escaped through the boundary, at 0.7 ps. Insets show enlarged trajectory of the test electron. (The magnetic field unit is $B_{unit}\approx10$ kT, the electric field unit is $E_{unit}\approx 3$TV/m.)}
\label{accel}
\end{figure}

The trajectory of one of these electrons is shown in Fig. \ref{accel}, on the background of the averaged $z-$component of the magnetic field (panel (A)),  on the background of the averaged radial component (normal to the target surface) of the electric field (panel (B)), and, in the subsequent panels, on the background of the not averaged $z-$component of the magnetic field for selected regions (panels (C) and (D)). It can be seen from the enlarged insets with the electron trajectory (panels (C) and (D)), that the initial energy gain is stochastic.  Eventually, the electron escapes the confining fields, crosses the target with the gained energy, and after that it is moving as a free particle. 

\begin{figure}{\begin{center}\includegraphics[width=0.4\textwidth]
{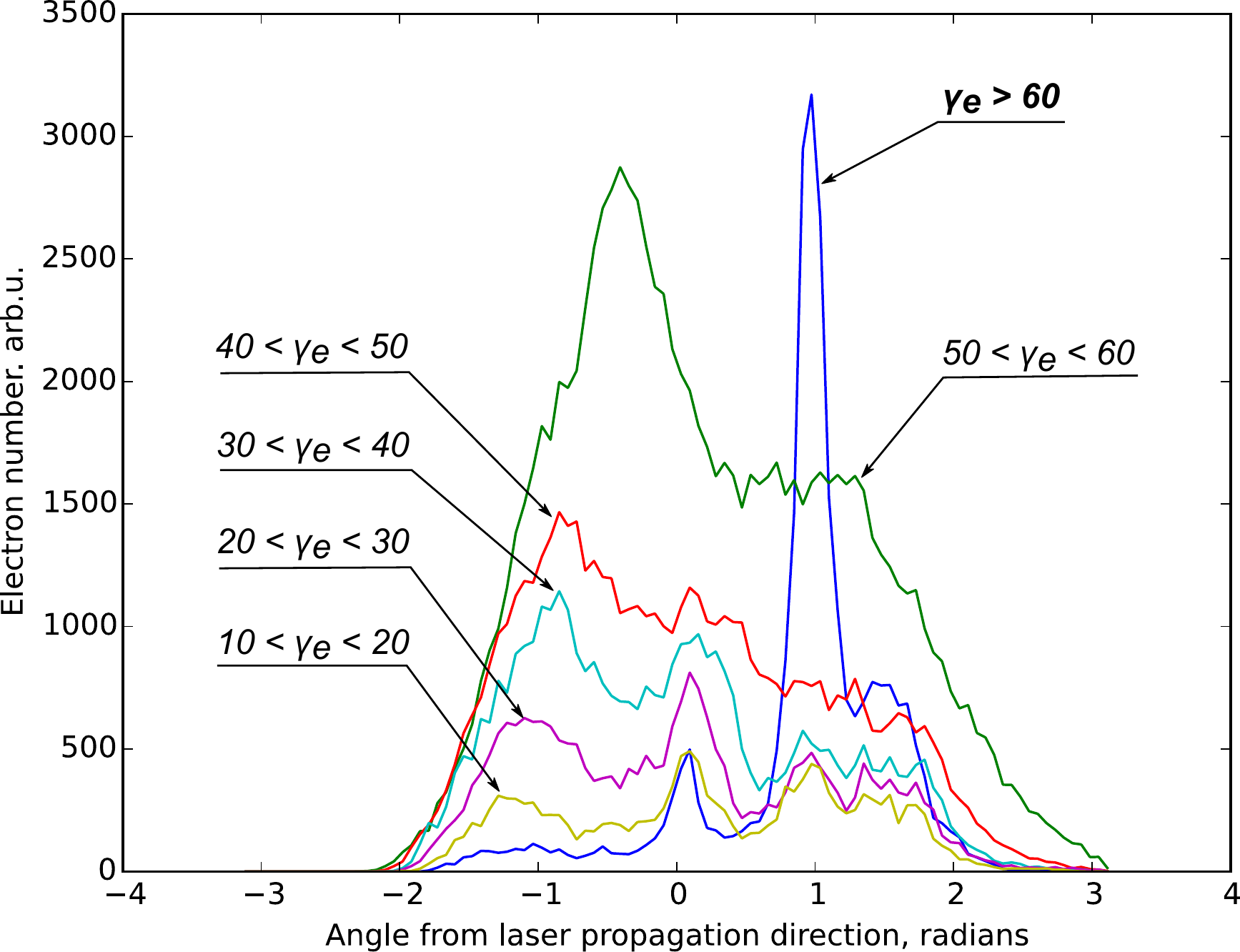}\end{center}}
\caption{Angular distributions for the accelerated electrons for different energy ranges. Zero angle corresponds to the propagation direction of the laser pulse.}
\label{ang_distr}
\end{figure}

While electrons may go through the target, the laser beam follows the surface, so the target curvature allows controllable separation of the beam and the accelerated particles. The particles may keep their energy, when the interaction conditions abruptly change and the adiabaticity is violated. This is similar to the case where a short laser pulse is reflected from a thin solid target at normal incidence, as described in \cite{Glazyrin-qe15}. In the front part of the laser pulse the electrons are accelerated and move generally with the laser. For the highly accelerated electrons, where $R_e\gg R_s$, the pitch angle approaches the angle of grazing incidence, so these electrons approach the target surface at the angle $\theta_0\approx \theta_{gr}$. At this stage the acceleration physics reminds the one presented in Ref.\cite{Pukhov-pop99}, though the origin and geometry of the quasistatic fields are significantly different. The fastest electrons, selected by the geometry constrains, form a beam with a narrow angular distribution. Figure \ref{ang_distr} shows the angular distribution of electrons in different energy ranges. A broad angular distribution at low energies transforms into a very narrow peak for electrons with $\gamma_e>60$.  For a grazing incidence of a laser pulse on a flat target the similar effect was reported in \cite{Rao-apl15}. The energy scaling, $\gamma_{e}^{(\max)} \sim 2 a_0 (R_s \omega/c)\sin\theta_{gr}$, corresponds to a continuous ``$\vecj \times \vecB$'' electron acceleration as it propagates through the plasma layer to the surface, and gives $\gamma_{e}^{(\max)}\approx 250$ for $R_s=100 ~\mu$m, $a_0=3$, $\theta_{gr}\approx 25\degree$. Note that with a focused beam, $\gamma_{e}^{(\max)}$ would decrease by a factor which depends on the exact interaction geometry due to the intensity variation along the acceleration path.   


The obtained energy scaling exceeds substantially the usual scaling for the given parameters $\gamma_e\gg \gamma_{e}^{(pond)}\equiv{1+a^2_0/2}\lesssim10$, (see e.g. \cite{Hartemann-pre95, Arefiev-pop14}).
The acceleration takes place for the electrons which are injected into the laser field with the apporpriate initial phase. Various mechanisms of electron injection have been discussed in the literature: applied magnetic or electric fields \cite{Moore-pre99, Arefiev-pop14, Buts-pasa04, Kemp-pre09, Paradkar-pop12}, colliding laser pulses  \cite{Sheng-prl02, Bulanov-jpp17, Gong-pre17, Bourdier-physd05, patin-lpb06}, superposition of traveling transverse and longitudinal  waves \cite{Bochkarev-ppr14} or a resonant excitation of surface plasma waves on a periodic surface structure \cite{Raynaud-pop07, Riconda-pop15}. 
Here, spontaneous magnetization may boost electron acceleration if it exceeds certain threshold value, similarly to the results in Refs. \cite{Moore-pre99, Buts-pasa04}. To unravel the injection mechanism, we performed a separate 3D modeling of the motion of test electrons in the given laser field with imposed constant magnetic field, with geometrical constraints simulating the target boundary. The stochasticity analysis of their motion was performed with the conventional methods \cite{liberman-lihtenberg-eng, Bochkarev-ppr14}.
%
%
%
The field parameters were the same as in PIC simulations. They were defined by the $y$-component of the laser vector potential, $ A^{\rm L}_y(\xi\equiv x-ct)= E_0 c /\omega_0 \exp(-\xi^2/\tau_L^2)\times\exp(-(y^2+z^2)/w^2_{0})\sin(\xi \omega/c )$, and a constant magnetic field $\vecB_{z0}$ directed along $z$ axis, so that the total vector potential was  ${{\vecA}}=\{-y B_{z0}/2, x B_{z0}/2+ A^{\rm L}_y,0\}$. 
The trajectories of ten thousand test particles were calculated using the equations
$
\frac{d}{d t}\left({\vecp}-{e{\vecA}/c}\right)=
-{e}/{c}\cdot\left({\bf v}{\bf \nabla} {{\vecA}}  +{\bf v}\times{\rm rot}\,{\bf   A}\right), ~ \dot{\bf r} = {\bf v},
$
${\vecp}=m_e \gamma_e {\bf v}\,$,
%
%
%
with initial positions distributed equally spaced in 
the simulation zone, which had a shape of a cylindrical section, with a width of $16~\mu$m $\times$ $16 ~\mu$m in the plane $(y,z)$ transverse to the laser propagation direction. In the plane $(x,y)$ the simulation zone was restricted by an arc with the radius $R_s=100~\mu$m with the center at $x=0$, $y=R_s$ and by straight lines $x=0$ and $y=D=16~\mu$m. All boundaries were transparent for the particles.

\begin{figure}[!ht]
\includegraphics[width=0.4\textwidth]{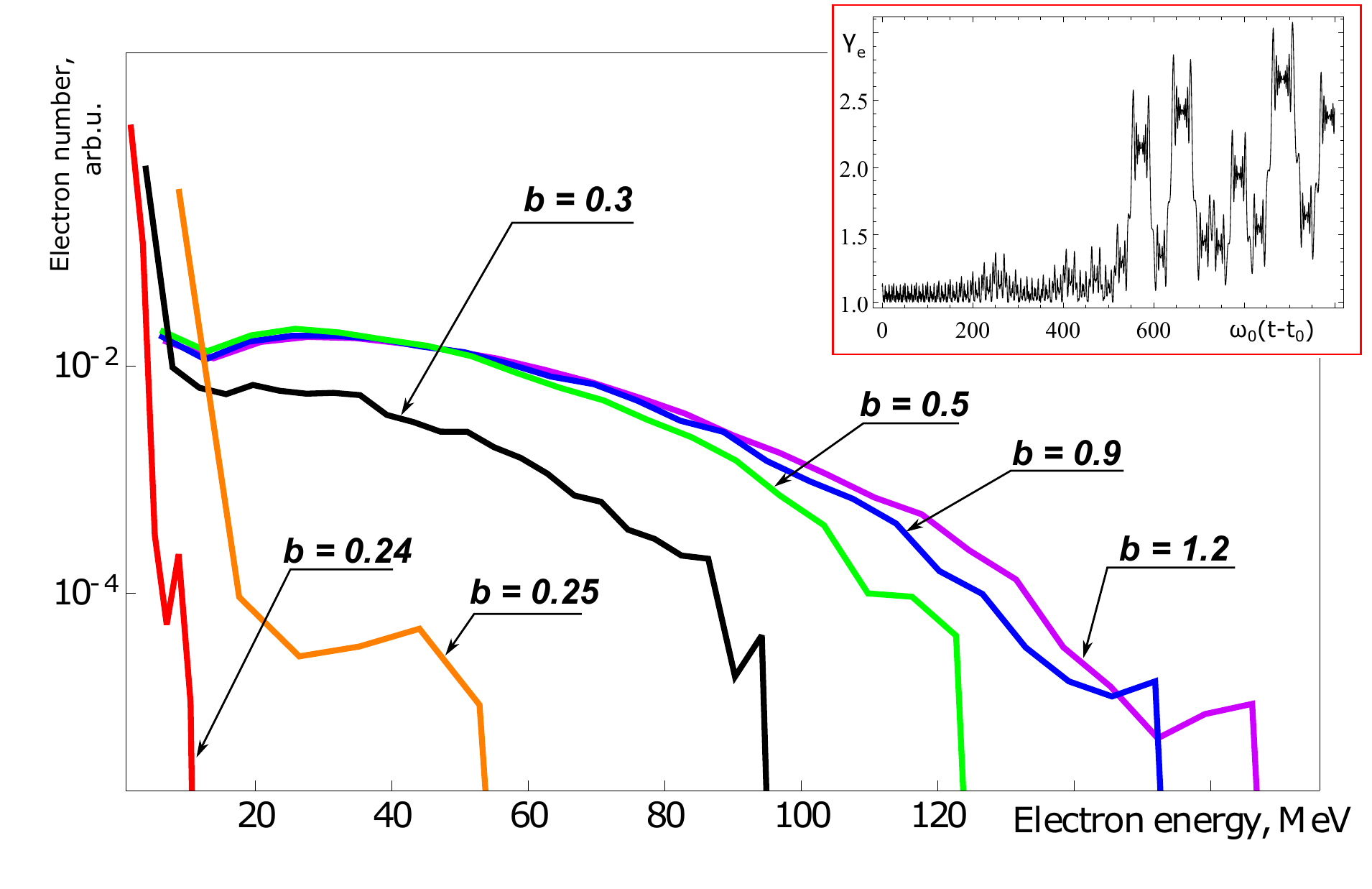}
\caption{Electron energy distributions at the end of simulation for different values of the imposed magnetic field, obtained from the test-particle modeling in given fields. Inset shows an abrupt electron energy increase at some time moments, when resonant conditions for the particle acceleration are achieved.}
\label{test-part-spectra}
\end{figure}

It was found in the test-particle approach, that for efficient acceleration, the magnetic field amplitude should exceed a threshold value, $  b\equiv e B_{z0}/m_e \omega_0c \geqslant b_{th}\approx0.25$. There is almost no energy gain below the threshold, while above the threshold the acceleration becomes very efficient, and electron energy increases abruptly to $\gamma_e\gg \gamma_{e}^{(pond)}$, see electron energy distributions for the different magnetic field amplitudes in Fig. \ref{test-part-spectra}. 
The threshold value $b_{th}$ is defined by the resonant condition between the amplitude of an electron quiver motion $\approx a_0c/\omega_{0}$ \cite{Hartemann-pre95} and the Larmor radius $m_e c/eB_{z0}$, so that the threshold may be estimated as $b_{th}\approx 1/a_0$, being consistent with the numerical value for $a_0\approx 3$.


The acceleration mechanism comprises two steps. First, a particle is randomly walking under the action of the laser field and the quasistatic fields, up to the moment when a resonance occurs, then the particle enters in the second phase of the direct laser acceleration. Unlike the relativistic betatron resonance \cite{Pukhov-pop98}, in our case the resonance takes place when electron energy is small, see the inset in Fig. \ref{test-part-spectra}. Depending on the conditions at the moment of injection, the particle may be then efficiently accelerated up to the target wall and escape, or it may be thrown away from the region of acceleration. 

Our theoretical findings are supported by the observation of the efficient electron acceleration by a laser field on a curved metallic surface.
The experiment was performed at the LFEX laser facility \cite{Miyanaga-jdp06} in the Institute of Laser Engineering (ILE) of the Osaka University.
The target had a shape of a cylindrical bent foil with a curvature of 250 $\mu$m (diameter of 500 $\mu$m) at $0\degree$ decreasing down to 200 $\mu$m at $360\degree$, so having a 100 $\mu$m-width slit as an entrance for the laser beams. The cylinder length was 500 $\mu$m and the foil thickness was 10 $\mu$m.
Three beams, called H1, H2 and H4, of the LFEX laser were tightly focused on the inner surface of the curved target through the entrance slit.
The spot diameter of overlapped laser beams was 60 $\mu$m containing 50\% of the laser energy.
The total laser energy and pulse duration were 710 $\pm$ 10 J and 1.3 $\pm$ 0.5 ps at full-width-at-half-maximum, so the laser intensity on the target was 9.7 $\times$ 10$^{18}$ W/cm$^2$.
The contrast between the main pulse intensity and the nano-second foot pulse was better than 10$^{9}$ \cite{Fujioka-pp16}.

Energy distribution of electrons emanated from a curved target into vacuum was measured with two electron spectrometers (ESM).
The first one (ESM1) was located on the laser incidence axis being referred to as $0\degree$. Another one (ESM2) was located at $30\degree$ off from the laser incidence axis.
ESM2 captures electrons accelerated along the curved surface by the first reflected laser light according to a simple geometrical consideration.
The ESM consist of two magnet poles aligned parallel to each other.
The spatial profile of the magnetic field was measured by a sensor and the average field strength was 0.7 T.
Electron spectra were recorded with image plates.
Energy dispersion and response of the ESM systems were calibrated by using an electron beam generated at the L-band LINAC facility \cite{Ozaki-pfr10}.

\begin{figure}
	\begin{center}
		\includegraphics*[width=0.45\textwidth]{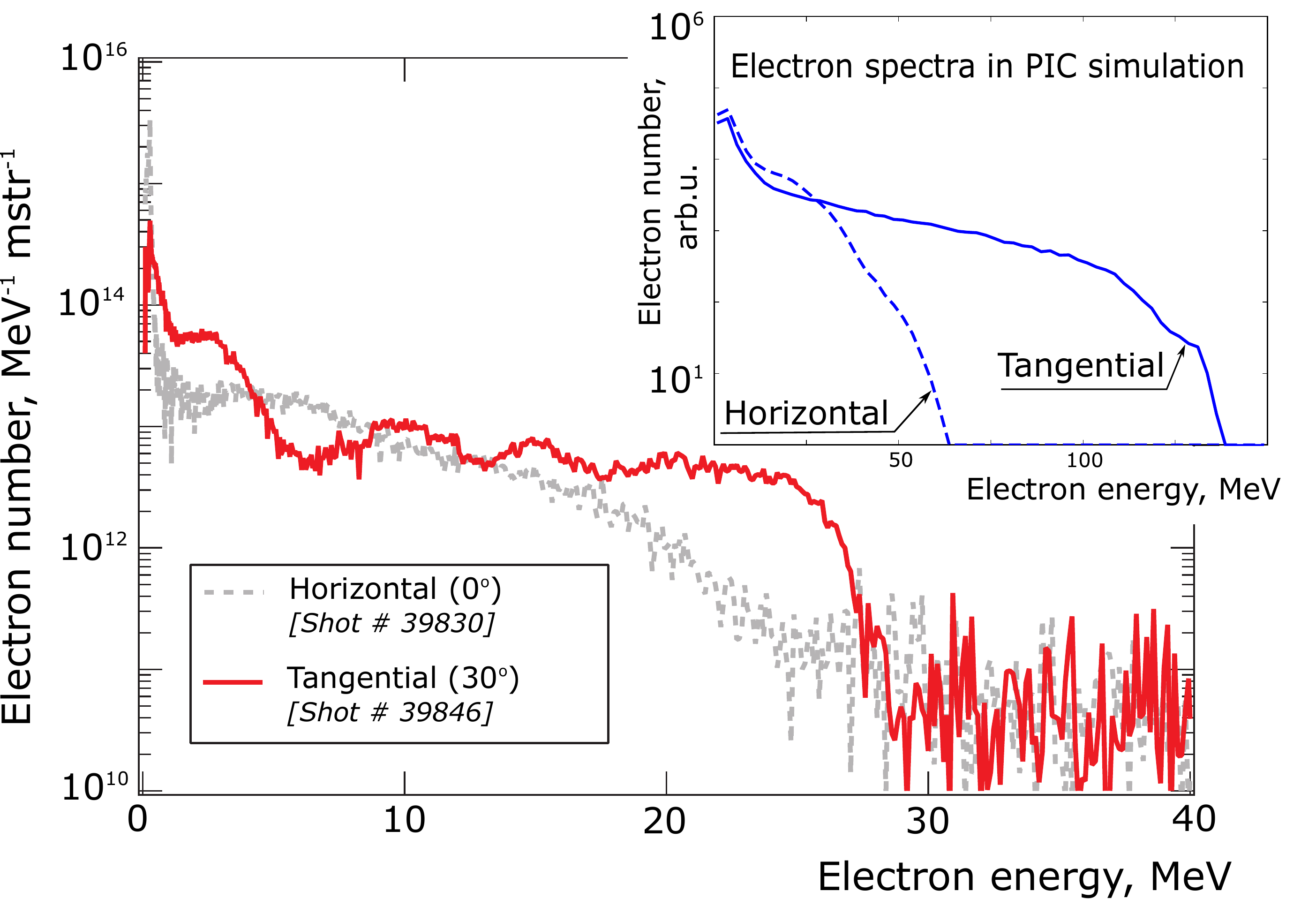}
	\end{center}
	\caption{Electron energy distribution measured along the laser incident axis (gray dot line) and at $30\degree$ from the laser incident axis (red solid line). \label{fig: elec_distribution_exp}}
\end{figure}

Figure \ref{fig: elec_distribution_exp} shows the energy distribution of accelerated electrons in two directions, obtained by ESM1 and ESM2.
The vertical axis shows the number of accelerated electrons per unit solid angle assuming isotropic distribution of emission ($4\pi$ srad).
The electron distribution can be approximated by an exponential function with the effective temperature of 6.5 MeV  in the range from 5 to 25 MeV at $0\degree$ (ESM1, gray dot line), which is a common figure for planar targets in high-intensity laser-plasma interaction experiments. In contrast, ESM2 at $30\degree$ that is closer to the direction of specular reflection of the incident laser beams from the curved surface, detected a quite different electron energy distribution (red solid line).
The observed plateau extends from 5 to 25 MeV, with a sharp cutoff at 27 MeV.

\begin{figure}
	\begin{center}
		\includegraphics*[width=0.45\textwidth]{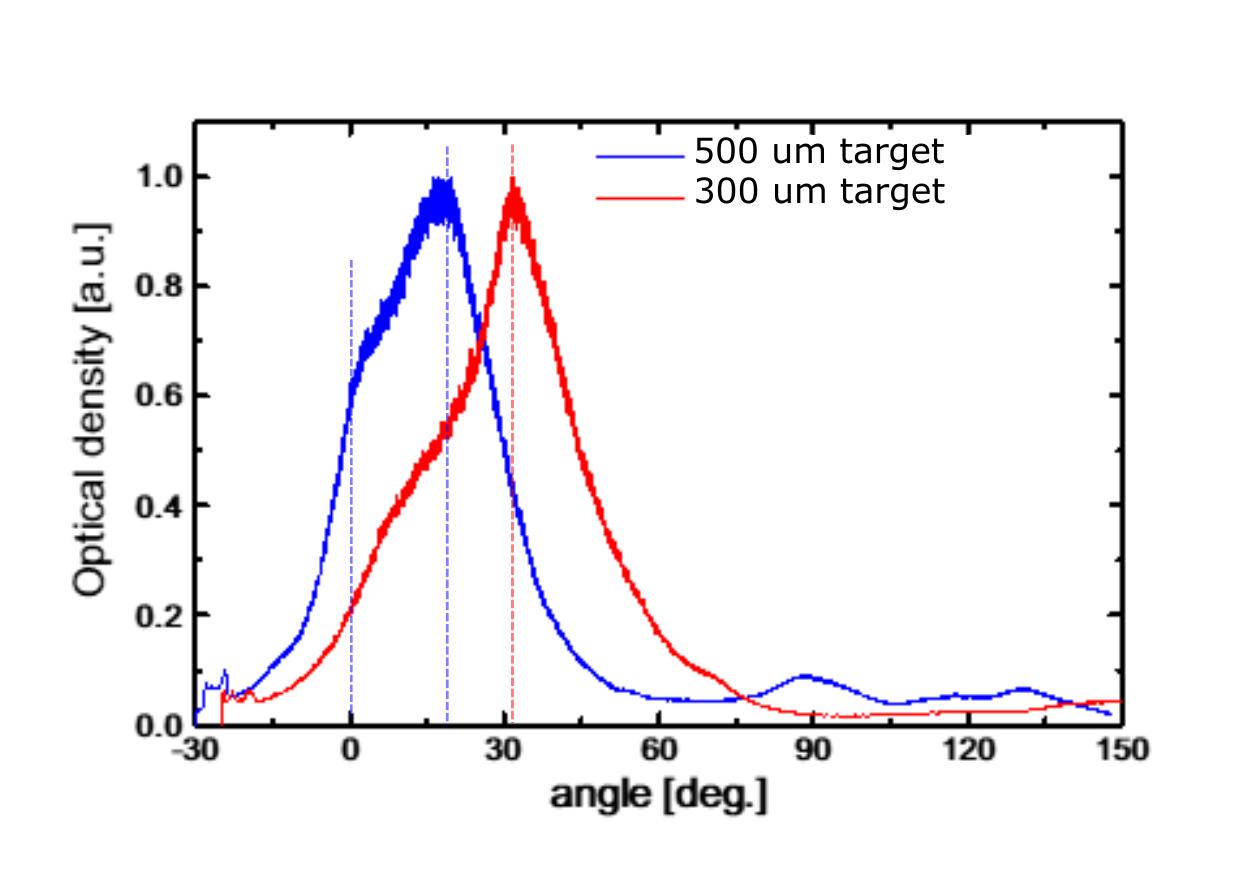}
	\end{center}
	\caption{Electron angular distribution obtained with radiochromic films. The ion signal is suppressed by filters. The blue and red curves correspond to the targets with diameters 500 and 300 $\mu$m respectively.\label{fig: e-ang}}
\end{figure}

Electron angular distribution was measured with radiochromic films, covering the angular range from $-30\degree$ to $180\degree$ from the laser beam direction in the plane perpendicular to the axis of cylinder. The obtained optical density is shown in Fig.~\ref{fig: e-ang}, where blue and red vertical lines indicate the tangent direction for the two targets with diameters 500 and 300 $\mu$m respectively. The maxima are observed in the tangent directions, according to the simple analysis presented above. 


%
%
%
%
%


In conclusion, we demonstrated a new method for electron acceleration with intense laser pulses.
Being simple in realization, it combines the known approaches for a controllable generation of spontaneous magnetic fields, electron guidance at grazing laser pulse incidence on solid-dense targets, and breakdown of adiabatic conditions at the end of acceleration phase. 
The main features of this acceleration method are (i) separation of the laser pulse and electron bunch and its collimation for the highest energies, (ii) generation of an electromagnetic guiding structure near the target surface providing a possibility for stochastic acceleration, and (iii) increasing the interaction distance by multiple laser beam reflection. The  angular distribution of accelerated electrons has a narrow peak for the highest part at an angle, which depends on the geometry of interaction. The characteristic plateau-like electron energy distribution is measured experimentally in the direction of laser specular reflection, as well as the angular distribution with a maximum at the tangent direction, confirming the theoretical ideas. 

\begin{acknowledgments}
The authors thank the technical support staff of ILE for assistance with the laser operation, target fabrication, plasma diagnostics, and computer simulations. 
The authors acknowledge Ms. Oshima for her efforts, especially on the target fabrication.
This work was supported by the Collaboration Research Program of ILE (2016Korneev and 2017Korneev); by the Japanese Ministry of Education, Science, Sports, and Culture (MEXT) and Japan Society for the Promotion of Science (JSPS) through Grants-in-Aid for Scientific Research (A) (No. JP16H02245); by Challenging Exploratory Research (No. JP16K13918); by the Bilateral Program for Supporting International Joint Research between Japan and Russia by JSPS.
This work has been carried out within the framework of the EUROfusion Consortium and has received funding from the Euratom research and training programme 2014-2018 under grant agreement No 633053. The views and opinions expressed herein do not necessarily reflect those of the European Commission.
A. Y. was supported by Grants-in-Aid for Fellows by JSPS (Grant No. 15J00850). The work was granted access to the HPC resources of CINES under allocations 2016-056129 and 2017-056129 made by GENCI (Grand Equipement National de Calcul Intensif). This work was supported by the Russian Foundation for Basic Research (\# 16-52-50019\,\hskip-9pt{\cyr{YaF}}), Excellence programm NRNU MEPhI. The numerical calculations were performed at Joint Supercomputer Center of the Russian Academy of Sciences and resources of NRNU MEPhI high-performance computing center.
\end{acknowledgments}

\bibliographystyle{new}
\bibliography{C:/Dropbox/BIBLIO/mendeley/library}

\end{document}